\documentclass[%
 reprint,
 amsmath,amssymb,
 aps,
]{revtex4-2}

\usepackage{graphicx}
\usepackage{dcolumn}
\usepackage{bm}
\usepackage{amsmath,amssymb}
\usepackage{braket}
\usepackage{mathtools}
\usepackage{float}

\usepackage{xcolor}

\begin{document}

\preprint{APS/123-QED}

\title{Revealing the embedded phase in single pixel quantum ghost imaging}

\author{Bereneice Sephton}
\affiliation{School of Physics, University of the Witwatersrand, Johannesburg, South Africa}
\author{Isaac Nape}
\affiliation{School of Physics, University of the Witwatersrand, Johannesburg, South Africa}
\author{Chan\'e Moodley}
\affiliation{School of Physics, University of the Witwatersrand, Johannesburg, South Africa}
\author{Jason Francis}
\affiliation{School of Physics, University of the Witwatersrand, Johannesburg, South Africa}
\author{Andrew Forbes}
\affiliation{School of Physics, University of the Witwatersrand, Johannesburg, South Africa}

\date{\today}
\begin{abstract}
\noindent We outline and experimentally demonstrate a method to image pure phase objects using traditional quantum ghost imaging with single pixel detectors. We provide a theoretical description of the process, showing how phase information is embedded in the correlation measurements of spatially entangled photon pairs in which only one of the pair interacts with the object. Using pairs of digital projective masks, one amplitude-only and one phase-only, we reconstruct two partial phase images of the object. Using both these images, we can retrieve the full detailed phase profile of the object without ambiguity, which we demonstrate on objects exhibiting phase steps and gradients.
\end{abstract}
\maketitle



In quantum ghost imaging, one photon from an entangled pair interacts with the object and is bucket-detected (no spatial resolution) while the other that does not interact with the object is directed to a spatially resolved imaging detector \cite{moreau2018ghost, tasca2013influence}. While neither photon alone has information of the object, an image can be formed by measuring the mutual correlations through coincidences. Initially ascribed to the quantum entanglement of photons \cite{Pittman1995,abouraddy2001role}, it has since been shown that ghost imaging is possible using classical correlations too \cite{bennink2002two,Gatti2004}. Ghost imaging has seen significant growth in the past two decades \cite{padgett2017introduction,edgar2019principles,gibson2020single,moreau2019imaging}, with the promise of enhanced resolution \cite{moreau2018resolution,toninelli2019resolution}, and expanding their application to the imaging of photosensitive objects by lower photon fluxes \cite{aspden2015photon,morris2015imaging}, as well as extending into the x-ray \cite{Pelliccia2016,Yu2016} and electron wavelength spectrum \cite{li2018electron,trimeche2020ion}. Imaging quality has steadily improved while acquisition times have steadily decreased, fueled by advances in computational \cite{Shapiro2008,Erkmen2010}, compressive sensing \cite{Katz2009,Zerom2011} and deep learning based methods \cite{Lyu2017,Shimobaba2018,Komuro2020,Rizvi2020,Rodriguez-Fajardo2020,Moodley2021}.

Quantum ghost imaging has traditionally been used to image amplitude objects, first with 2-dimensional spatial projective masks such as Hadamard \cite{Duarte2008} and random \cite{Shapiro2008}, and more recently with advanced camera systems.  Despite these advances, phase imaging has progressed much slower, with only limited advances to date. This has been achieved with enhanced phase contrast \cite{aspden2013heralded}, interferometric approaches \cite{Zhang2019,Ndagano2021}, more recently using quantum-correlation enabled Fourier ptychography \cite{aidukas2019phase} and polarisation enabled holography \cite{Defienne2021p} all using modern cameras. While these techniques can produce the desired results, they have some shortcomings: the phase contrast imaging techniques return partial information and work well for objects with binary phases, the ptychography and interferometric approaches are alignment intensive, prone to instability, and are costly since single photon cameras are required. Remarkably, digital holography, which is intrinsically an interferometric technique, does not require physical path interference, but rather intelligently constructed masks, therefore enabling one to produce the desired interference effects with the aid of diffraction. Therefore, devising phase inference strategies employing digital masks that can be encoded on inexpensive spatial light modulators which are easy to align in quantum experiments and readily available in most optical labs, make this approach a viable solution.

In this work we show that this required interference can naturally arise in the correlation measurements used in traditional quantum ghost imaging employing binary phase digital holograms, avalanche photo-diodes and a coincidence counting device. By adjusting the conventional 2D spatial projective masks applied to the reference photons, we project their entangled counterparts onto a superposition of themselves. This interference effect eliminates the need for direct interferometry and sensitive high resolution cameras. As a result, we show that conventional single-pixel ghost imaging setups are already able to see detailed phase objects without any physical adjustments, but rather only require an additional spatial mask projection that is readily made from the masks already used. 

\begin{figure}[ht]

\includegraphics[width=\columnwidth]{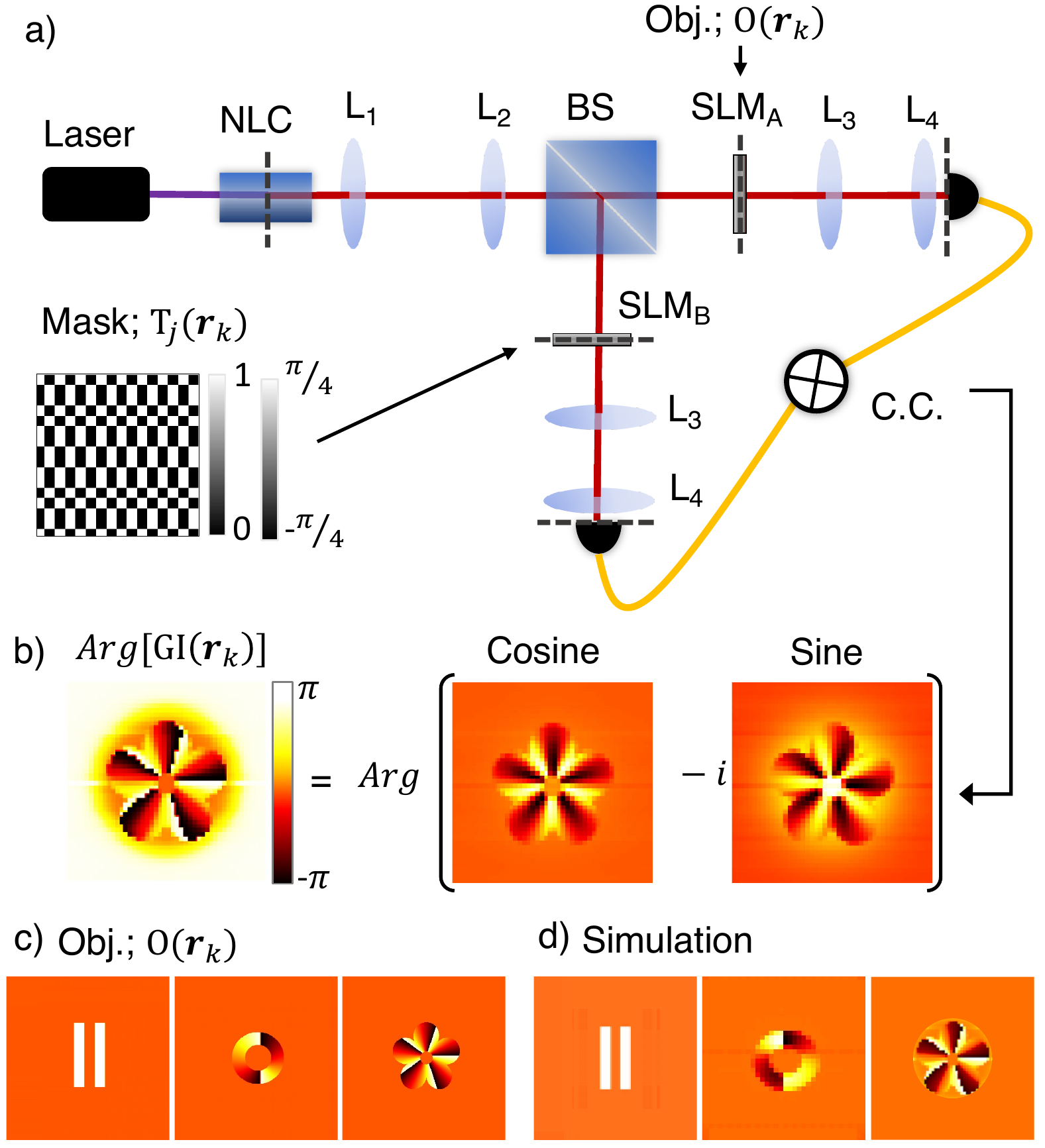}
\caption{\label{fig:exp} A schematic diagram of the implemented quantum ghost imaging setup. (a) Entangled bi-photons are produced via spontaneous parametric down-conversion (SPDC) at the NLC. The entangled photons are spatially separated and each imaged onto a SLM. Required holograms are displayed on each SLM. Coincidence measurements are done between both paths and the coincidences are used to reconstruct an image of the object. For each object two measurements are taken, as numerically simulated in (b). These two images are then combined such that the argument reveals the total phase. (c) Shows the digital phase-only objects used in the experiment, while the simulated image reconstructions are shown in (d) showing how the digital objects are expected to perform. L$_1$ = 50 mm, L$_2$ = 750 mm, L$_3$ = 750 mm, L$_4$ = 2 mm.} 
\centering
\label{fig:setup}
\end{figure}

To illustrate this, we implemented an all-digital ghost imaging setup with spatial masks as shown schematically in Fig. \ref{fig:exp} (a). Light from a diode laser (wavelength $\lambda$ = 405 nm) was used to pump a temperature controlled type I PPKTP non-linear crystal (NLC). The temperature was set to optimally obtain collinear SPDC of entangled bi-photons at a wavelength of 810 nm. Any unconverted pump photons were then filtered out with a bandpass filter placed after the NLC. Next, the entangled photons were separated into two paths with a 50:50 beam splitter (BS) with the digital object encoded on a spatial light modulator (SLM) in path A (object arm), projective digital masks for imaging, also written to a SLM, in path B (image arm) with a blazed grating on each SLM. The object and image photons were coupled into single mode fibres connected to avalanche photo-diodes (APDs) for detection in coincidence (C.C.). The SLMs and fibres were all positioned in the image plane of the crystal as indicated by dashed lines in the figure.

Here we elucidated the ``hidden'' phase features of the objects from correlation measurements between the photons in each arm. In the discrete position basis, $\{\ket{\textbf{r}_{j}}\}$, the (entangled) bi-photon state have correlations described by the state
\begin{equation}
    \ket{\Phi} = \frac{1}{N} \sum_{j=0}^{N-1} \ket{\textbf{r}_{j}}_A\ket{\textbf{r}_{j}}_B,
\end{equation}
\noindent where $N$ indicates the number of discrete position basis states that are correlated in the x or y coordinate. For most ghost imaging experiments, an object is decomposed using a complete set of projection states $\ket{M_{j}}$ or equivalently $\textbf{M}_{j}$ in matrix form.  We can thus describe the object in this basis following the decomposition 
\begin{equation}
\ket{O} = \sum_{j=0}^{N-1}\sqrt{p_{j}}e^{i\alpha_j}\ket{M_{j}}.
\label{eq:objectDecomp}
\end{equation}
Here $p_{j}$ is the probability of projecting onto the basis state $\ket{M_{j}}$ and $\alpha_{j}$ are the corresponding phases. Furthermore, the spatial profile of the object is given by, $\braket{ \textbf{r}_{j} | O} = O(\textbf{r}_{j})$ corresponding to the complex matrix $\textbf{O}$. 

As an example, we will consider the basis modes constructed from the Walsh-Hadamard transform \cite{walsh1923closed} with column vectors $h_{n}$. In matrix form, the projection states are $\textbf{M}_{j\rightarrow (nm)} =  h_{n}\otimes h_{m}$ which comprises arrays of $\pm1$ pixels (in and out of phase).
After renormalisation following, $\ket{T_{j}} =1/\sqrt{2}\left( \ket{M_{0}} + \ket{M_{j}}\right)$, where $\braket{\textbf{r}_{j}| M_{0}} = 1/\sqrt{N}$, the final projection states correspond to matrices, $\textbf{T}_{j}^{\cos}$, that have entries that are 0s and 1s.

Traditionally the ghost image is reconstructed from the second order correlation function \cite{welsh2015near}
\begin{equation}
 \textbf{GI}_{\cos}= \frac{1}{N} \sum_{j=0}^{N-1}\Bigl[|c_{j}|^2- \braket{|c_{j}|^2}_N\Bigr] \textbf{M}_{j}.
 \label{eq:GIimages}
\end{equation}
The ensemble averages are computed as $\braket{A_j}_N = \sum_{j=0}^{N-1}A_j/N$. While it is common to use projection masks, $\textbf{T}^{\cos}_{j}$, in Eq.~(\ref{eq:GIimages}) in the reconstruction procedure, it suffices to use the Hadamard masks, $\textbf{M}_{j}$, instead  because $\textbf{GI}_{\cos}$ is independent of the reference mode, $M_{0}$ (see Supplementary Material).
To reveal the spatial phase information in the object, we first compute $\braket{|c_j|^2}_N = p_0/2 + \braket{p_j}_N/2 + \braket{\sqrt{p_o p_j}\cos(\Delta \alpha_j )}_N$. After substitution into Eq.~(\ref{eq:GIimages}) and further simplification (see Supplementary Material), the measured ghost image is
\begin{align}
 \textbf{GI}_{\cos} =&  \frac{1}{N}\left( \sqrt{p_0} \text{Re} \left( \textbf{O} \right) +  \frac{1}{2} \textbf{H}_{d} \textbf{P} \textbf{H}_{d}^\dagger - g_{\cos} \textbf{R}_0 \right),
 \label{eq:GIcos}
\end{align}
where the first term is the real part of the object,  $\text{Re} \left[ \textbf{O} \right] = |\textbf{O}|\cos \left( \text{arg} \left( \textbf{O} -\alpha_0 \right) \right)$. The second term is the Hadamard transform of matrix, $\textbf{P} = |\textbf{H}_d \textbf{O} \textbf{H}_d ^ \dagger |^2$ where $|\cdot|^2$ is the element wise absolute value squared. Each "pixel" entry in $ \textbf{P} $ maps onto a probability $p_j$. The last term removes the DC component, by subtracting the first pixel $\textbf{R}_0 \rightarrow \delta_{j0}\ket{r_j}$ weighted by the factor $g_{\cos}$. Moreover, the weighting, $g_{\cos}$, is a constant that depends on the ensemble average of the probabilities,  $p_j$, and cosines of the phases (see Supplementary). It can be shown that in some cases the second term does not overlap spatially with the desired first term, and therefore  can be removed from the constructed image by simply cropping it out (see Supplementary).

Following the same arguments, the sine of the phase, corresponding to the ``imaginary part" of the ghost image can be computed by applying the same analysis but with an adjustment to the projection masks, i.e., the new projections are 
 $\textbf{T}_j{^{\sin}}= \frac{1}{\sqrt{2}} ( \textbf{M}_j + i\textbf{M}_0)$, having a relative phase of $\pi/2$. The resulting detection probabilities are then $|c_j|^2= p_0/2 + p_j + \sqrt{p_o p_j}\sin(\Delta \alpha_j )$. Accordingly, the detected ghost image becomes
\begin{align}
 \textbf{GI}_{\sin} =&   \frac{1}{N} \left( \sqrt{p_0} \text{Im} \left( \textbf{O} \right) +  \frac{1}{2} \textbf{H}_{d} \textbf{P} \textbf{H}_{d}^\dagger - g_{\sin} \textbf{R}_0 \right),
 \label{eq:GIimageSolved1}
\end{align}
having an embedded imaginary part of the object profile where  $\text{Im} \left( \textbf{O} \right) =|\textbf{O}|\sin \left( \text{arg} \left( \textbf{O} \right) -\alpha_0 \right)$. The constant $g_{\sin}$ weights the DC component to be subtracted and depends on the ensemble average of the probabilities,  $p_j$, as well as the sine of the phases (see Supplementary). Realisation of imaginary image reconstruction is then achieved using Eq.~\ref{eq:GIimages} where the projected sine probabilities $c_j$ are paired with the corresponding Hadamard masks $\textbf{M}_j$ such that a real image is formed with the required sine phase information.

Accordingly, two image reconstructions (cosine and sine) are required to fully reconstruct a phase object, as illustrated in Fig.~\ref{fig:setup} (b). Applying this  to three phase objects, shown in Fig.~\ref{fig:setup} (c), results in the simulated outcomes shown in Fig.~\ref{fig:setup} (d). Experimentally, the cosine projection was obtained by displaying masks on the SLM$_\text{B}$ in which the blazed grating was turned off (amplitude of 0) in certain pixels and on (amplitude of 1) in others. The location of the on or off pixels was determined by the specific mask in the basis set (for full details of the scanning technique see Refs. \cite{Rodriguez-Fajardo2020, Moodley2021}). The sine projection was then obtained by displaying the same masks with an identical pixel arrangement, however, the "on" and "off" pixel states were instead replaced with $1+i$ and $1-i$ phase states, respectively. 

\begin{figure}[ht]
\includegraphics[width=\columnwidth]{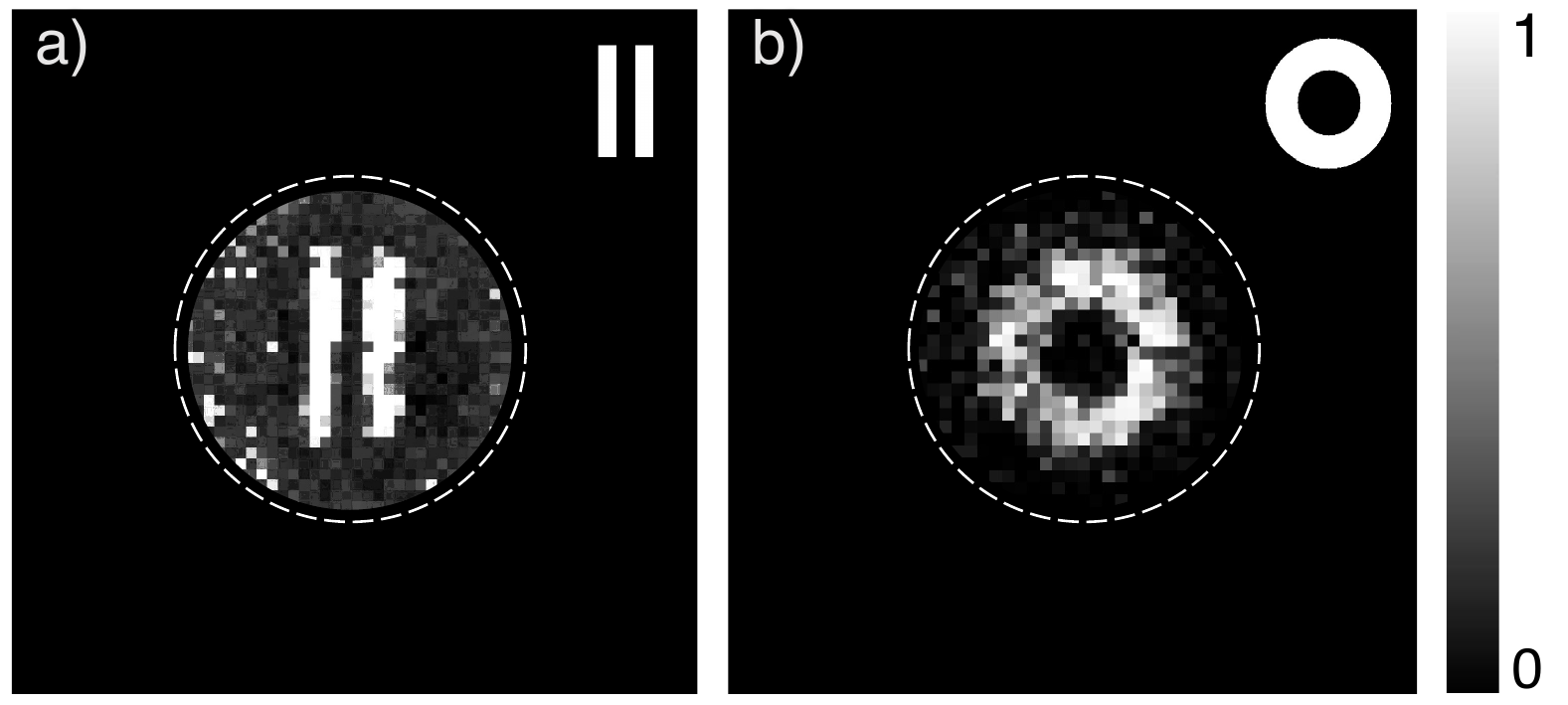}
\caption{\label{fig:amplitude} Reconstructed amplitude-only images using the Walsh-Hadamard masks for (a) an intensity slit, and (b) an intensity ring. The outer area of the dashed white circle indicates the region in which noise was suppressed due to lack of SPDC signal. The amplitude-only objects are shown as insets.}
\centering
\end{figure}

Initially, we show the experimental reconstruction of two amplitude-only objects in Fig.~\ref{fig:amplitude}, namely two vertical slits (a) and an annular ring (b). These were achieved by modulating only the amplitude of the photons with SLM$_A$ (leaving a flat phase) and using the cosine projection masks on SLM$_B$ to detect and thus reconstruct the amplitude object according to the algorithm in Eq.~\ref{eq:GIimages}. Insets in the corners of the figure show the transmission masks used. A good correlation between the object and detected distribution can thus be seen. 

\begin{figure*}[t]
\includegraphics[width=\linewidth]{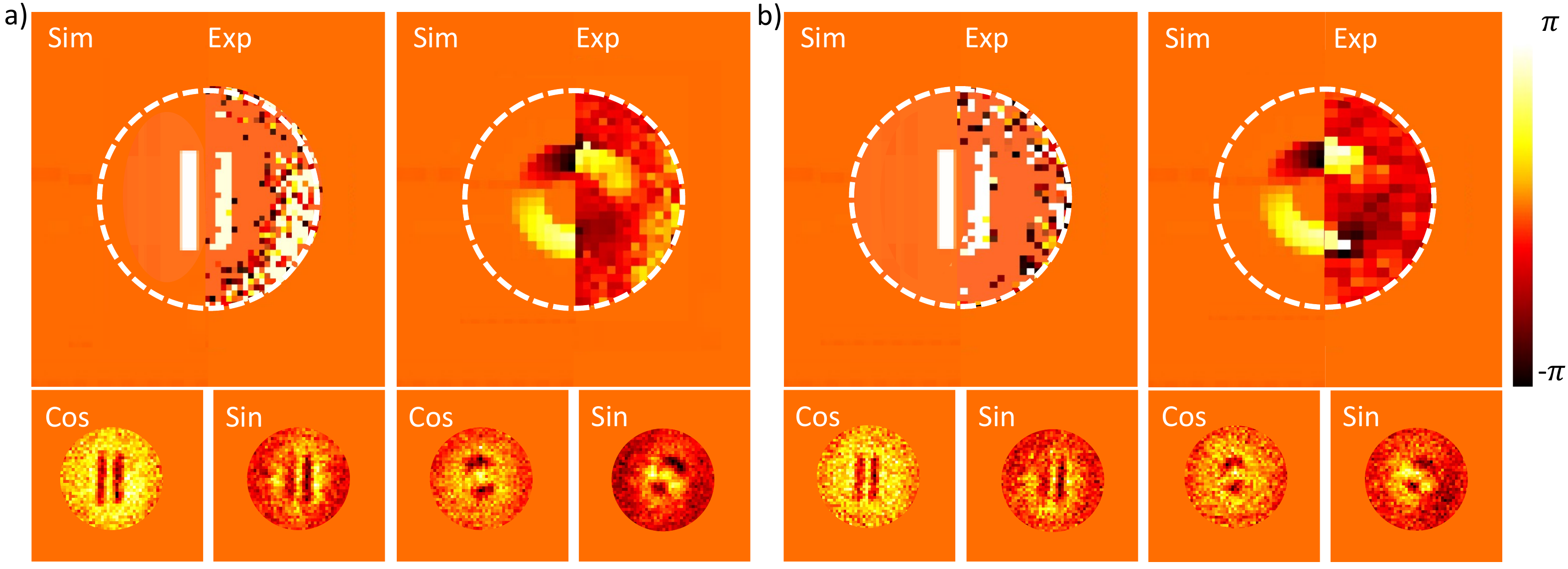}
\caption{\label{fig:comparison} Numerical simulations (Sim) showing excellent agreement with the Experimental reconstructions (Exp) for the $pi$-phase slit and the azimuthal gradient ring using (a) Walsh-Hadamard and (b) random masks. Experimental images were denoised with image processing tools and contrast adjustment after reconstruction. The outer area of the dashed white circle indicates the region in which noise was suppressed due to lack of SPDC signal. Insets show the corresponding cosine and sine components of the experimental reconstructions.}
\centering
\end{figure*}

By converting the amplitude objects in Fig. \ref{fig:amplitude} to pure phase objects shown in Fig \ref{fig:comparison}, we show that phase information is also embedded in the ghost imaging technique and retrieved by an additional sine projection. Here, we illustratively compare the numerical simulation (Sim) with the experimentally (Exp) reconstructed phase-only $\pi$-phase slit and the azimuthal phase ring. Insets at the bottom show both the experimentally reconstructed cosine and sine projections. While, previously, we focused on the Walsh-Hadadmard basis, here we also show that this phase retrieval can be extended to the pseudo-complete random basis. As can be seen, the total phase of the object is recovered for both the Walsh-Hadamard in (a) and random masks in (b), showing not only phase steps for the $\pi$-phase slit, but also contrast in the phase gradient by the spiral features with the azimuthal phase ring. Apart from noise distortion, we show good visual agreement between the numerical simulation and the reconstructed experimental images. It may also be noted that a flat circular phase background persists within the detected region on the SLMs that was illuminated with the SPDC photons. Past this region, a larger variation of noise persists due to lack of signal. We have therefore suppressed the noise in the region where there is no signal, illustrated by a dashed white circle.

\begin{figure}[ht]
\includegraphics[width=\columnwidth]{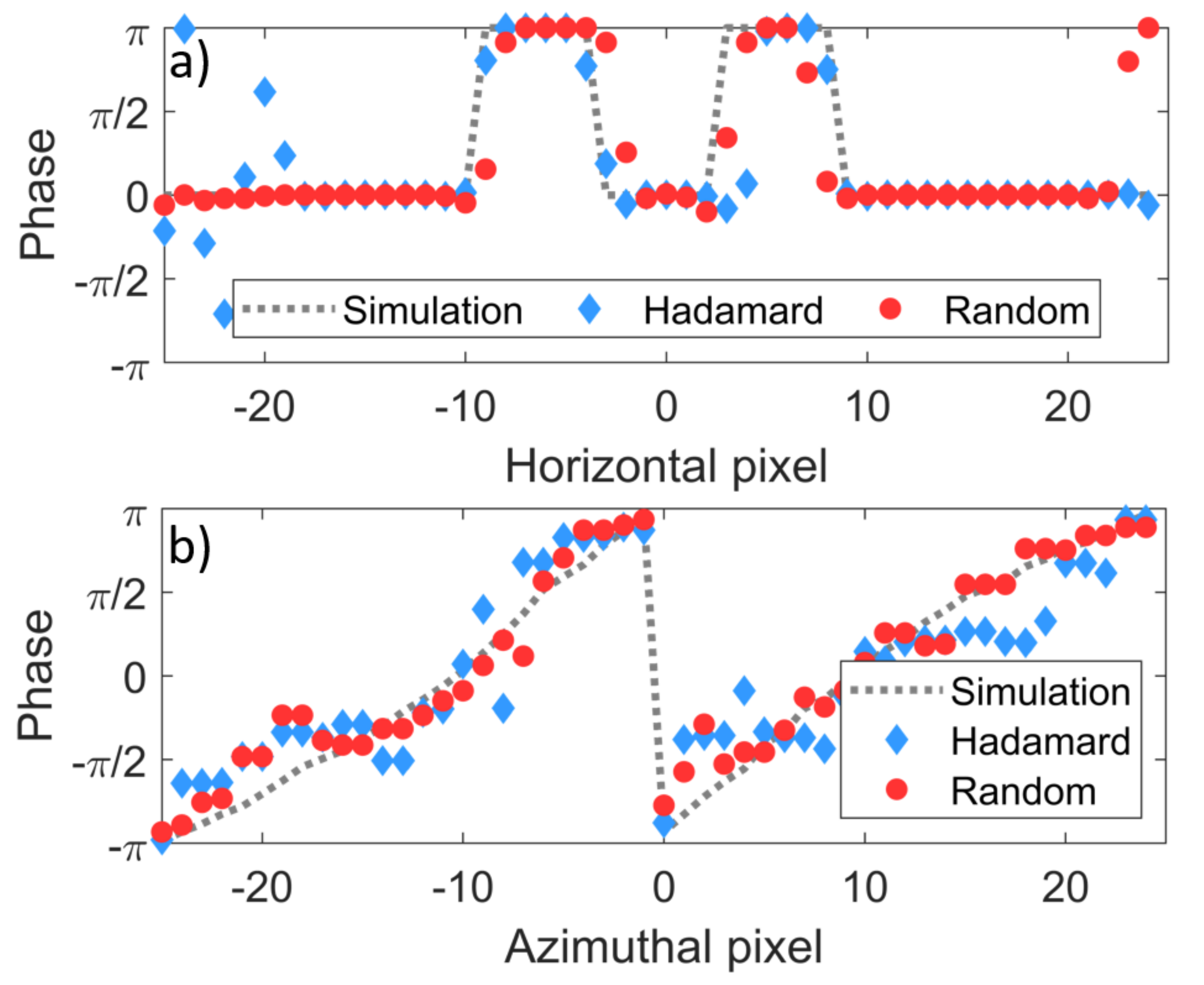}
\caption{Phase image cross-sections showing the phase value per pixel for the numerically simulated reconstruction given by the grey dotted line and experimental reconstructions for both the Walsh-Hadamard (blue diamonds) and random masks (red dots) for (a) the $\pi$-phase slit in the horizontal direction, and (b) phase ring in the azimuth direction for a radius set inside the ring.}
\centering
\label{fig:CrossSection}
\end{figure}

To assess the degree of agreement between the numerical simulations and experimental phase reconstructions in Fig.~\ref{fig:comparison}, we show cross-section plots in Fig.~\ref{fig:CrossSection}. The phase values per pixel of the slits are shown in (a) for a representative horizontal cross-section. We show good agreement between the numerical simulation (gray dotted line) and the experimental images for both the Walsh-Hadamard (blue diamonds) and random (red circles) basis reconstructions. Similarly, in Fig.~\ref{fig:CrossSection} (b), the phase values per pixel are given in the azimuth direction for a set radius within the annular ring. Again, we show excellent agreement between the numerical simulation and experimental reconstructions for both scanning methods. 

\begin{figure}[ht]
\includegraphics[width=\columnwidth]{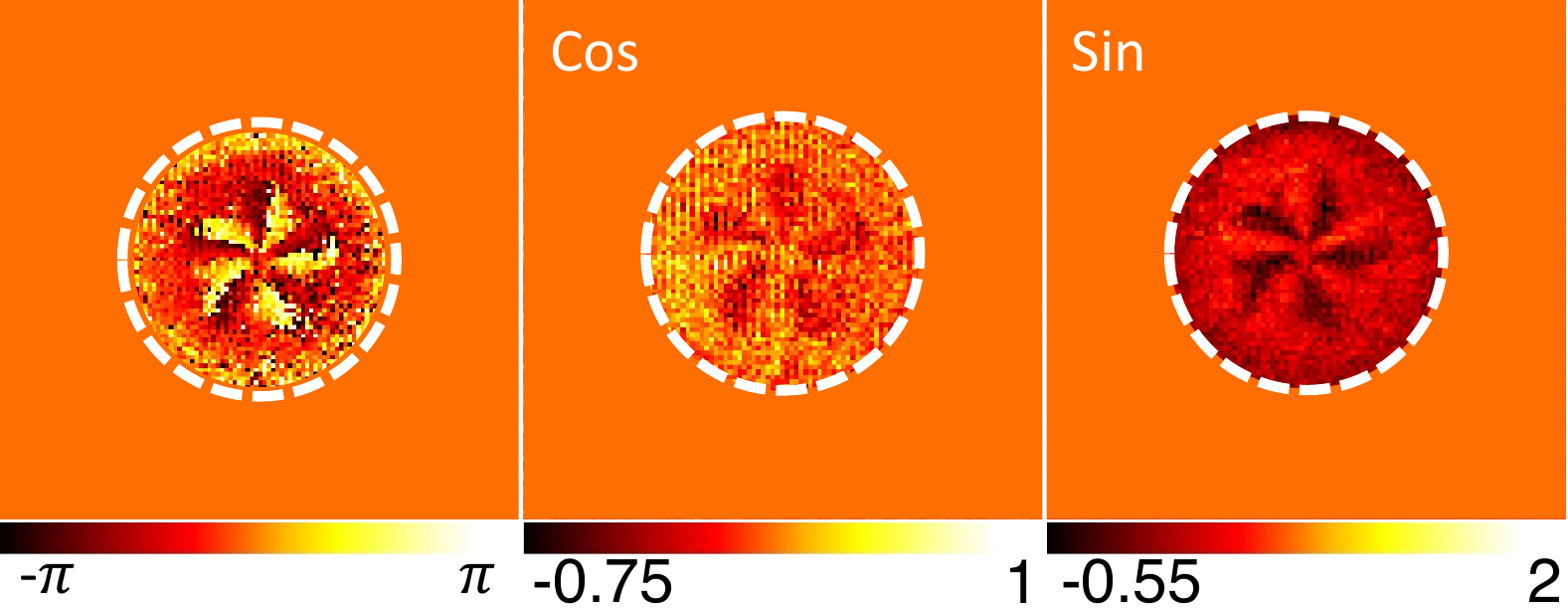} 
\caption{Experimental ghost image of a pure phase gradient spiral flower using Walsh-Hadamard masks (left). The full phase profile was denoised after reconstruction by image processing tools and contrast adjustment. The cosine (middle) and sine (right) components are shown for completeness.}
\centering
\label{fig:PhaseRecon}
\end{figure}

Lastly, in Fig.~\ref{fig:PhaseRecon} (a) we show the experimental reconstruction for the detailed spiral phase flower, which was shown numerically in Fig.~\ref{fig:setup} (b). Here we show both the cosine and sine projections (middle and right) which were used to reconstruct the full phase profile (left). We imaged at a high resolution of 128 $\times$ 128 pixels and used the Walsh-Hadamard masks. Importantly, we see that the use of 2D spatial projective masks reveals the entire phase structure including all phase steps and phase gradients albeit with the presence of noise. 

In contrast to other approaches,  we have reduced the number of necessary measured variables to two (cosine and sine), we require $n^2$ measurements for each variable, where $n \times n$ pixels is the image resolution. We showed that the required interference naturally arises from the correlation measurements without interferometry. We have, therefore, developed and implemented a stable, cost-effective quantum ghost imaging technique to reconstruct the full phase profile of a phase object by measuring a fewer number of variables while retrieving phase steps and phase gradients with the use of single pixel detectors and no need for computationally intense iterative algorithms to extract the phase.

In conclusion, we have shown that the 2D spatial masks used in traditional ghost imaging hold sufficient information to fully reconstruct and reveal complex phase structures. Here we demonstrated clear phase steps and azimuthal phase gradients within annular phase steps in the image reconstructions. Our cosine and sine projective masks were shown to reconstruct both counterparts of the image, which when combined, revealed the entire phase profile of our encoded phase-only objects. By comparing the phase values per pixel in the horizontal and azimuth directions between simulated and experimental reconstructions, we have shown good agreement for both the Walsh-Hadamard and random basis reconstructions, albeit with noise distortions present. The interference which is usually required to image a phase object is seen to be naturally present in the correlation measurements fundamental to traditional quantum ghost imaging with 2D spatial masks. We speculate that this fact has been overlooked due to the swift transition to sophisticated cameras.  Nevertheless, this all-digital set-up allows full amplitude and phase retrieval without complicated interferometric or computational techniques, and can be enhanced further by a judicious choice of mask type and image processing tools. This technique is well suited to be extended to complex amplitude object imaging and can be applied in both the physical and biomedical domains.


\newpage \clearpage

\onecolumngrid
\begin{center}
    \textbf{\Large Supplementary Information: Revealing the embedded phase in quantum ghost imaging}
\end{center}

\vspace{0.5 cm}

\twocolumngrid

\section{Revealing the embedded phase information}

Suppose we have an object that can be represented using a discrete pixel location states $\{ \ket{\textbf{r}_{j}}\}$. Since the object can be represented as a matrix, $\textbf{O}$, we can also use a matrix basis, $\{ \textbf{R}_j, \ j = 0,1, .. N-1 \}$ where, to map each pixel location in two-dimensional space. The matrices have $d$ by $d$ dimensions for corresponding x and y directions, respectively.

The quantity computed in the ghost imaging experiment can be written as \cite{welsh2015near}
\begin{equation}
 \textbf{GI}= 1/N \sum_{j=0}^{N-1}\Bigl[|c_{j}|^2- \braket{|c_{j}|^2}_N\Bigr] \textbf{T}_{j},
 \label{eq:GI}
\end{equation}
The coefficients $c_j$ correspond to the overlap probability between the object and the mask $\textbf{T}_{j}$ represented by a $(d,d)$ matrix. The masks, $\textbf{T}_{j}$, are computed from the Hadamard basis $\{\textbf{M}_j \equiv \ket{M_j}, \ j = 0,1,..N=d^2 \}$
\begin{equation}
 \textbf{T}_j= \frac{1}{\sqrt{2}} ( \textbf{M}_j + \textbf{M}_0).
 \label{eq:MaskEquation}
\end{equation}
 Each mode, $\textbf{M}_j$, is found by computing outer products of column vectors of the $d$ dimensional Walsh-Hadamard transform matrix $H_d$, i.e., $\textbf{M}_{j\rightarrow(m,n)} = h_n \otimes h_m$ and can be written as a superposition of the original pixel location basis, $\{ \textbf{R}_j, \ j = 1,2, .. d^2 \}$. Here $\textbf{M}_0$ is a reference mode that rescales the pixels (equivalently matrix entries) from having values proportion to -1 and 1 to now having 0s and 1s.

\begin{figure*}[t]
\includegraphics[width=\linewidth]{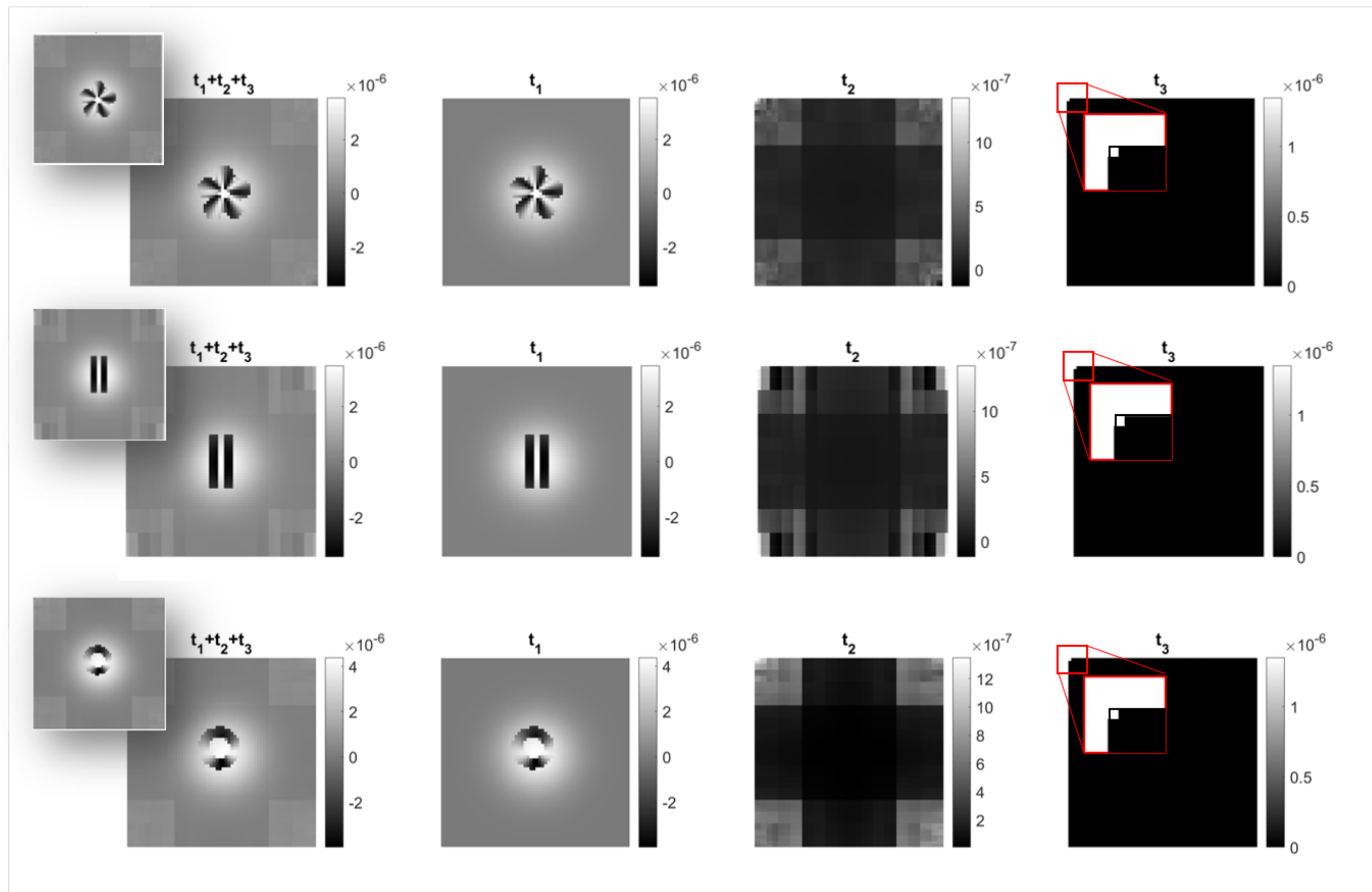}
\caption{Examples of the computed terms in the ghost imaging protocol. Here the labels $t_{1,2,3}$ correspond to the first, second and third term in $\textbf{GI}_{\cos}$. The inset in the firs column corresponds to the ghost image obtained from the algorithm in Eq~\ref{eq:GI}. The insets in the fourth column show the DC component to be subtracted.}
\centering
\label{fig:GIterms}
\end{figure*}

Given Eq.~\ref{eq:MaskEquation} and, we can express the overlap probabilities as
\begin{equation}
 \textbf{GI}= 1/N \sum_{j=0}^{N-1}\Bigl[|c_{j}|^2- \braket{|c_{j}|^2}_N\Bigr] \textbf{M}_{j},
 \label{eq:GIimage}
\end{equation}
showing that we only need the basis $\{\textbf{M}_j\}$ in the reconstruction. Next, we expand the detection probabilities
\begin{equation}
    \begin{split}
    |c_j|^2 &= |\braket{T_j|O}|^2,\\
    &= p_0/2 + p_j/2 + \sqrt{p_o p_j}\cos(\Delta \alpha_j),
    \end{split}
    \label{eq:decomProb}
\end{equation}
where $\Delta \alpha_j =  \alpha_j + \alpha_0$. The resulting probabilities are proportional to the joint probability between the two photons when measured in coincidence. Here we have assumed that the object can be decomposed as in the Hadamard basis following
\begin{equation}
 \textbf{O}= \sum_{k}^{N} \sqrt{p_k} e^{i\alpha_k} \textbf{M}_j,
 \label{eq:ObjMatrix}
\end{equation}
or equivalently using braket notation
\begin{equation}
 \ket{O}=  \sum_{k} \sqrt{p_k} e^{i\alpha_k} \ket{M_j},
 \label{eq:ObjState}
\end{equation}
with given probabilities $p_k = |\braket{O,M_K}|^2$ and phase $\alpha_k$.
Another important point to note is that the probabilities, $p_k$, can also be computed as a matrix, 
\begin{align}
 \textbf{P} &= \left( \sum_{j=0}^{N-1} p_j  \textbf{R}_{j} \right),
 \label{eq:Probmatrix}
\end{align}
which will be useful later for computing the final ghost image.

Next, we expand the coefficients in Eq.~\ref{eq:GIimage} using Eq.~\ref{eq:decomProb} and further simply it
\begin{align}
 \textbf{GI}_{\cos} =& 1/{N} \sum_{j=0}^{N-1}\Bigl[ \sqrt{p_0 p_j}\cos(\Delta \alpha_j) + \frac{p_j}{2} - \nonumber \\ 
             & \left( \braket{p_j}_N/2 + \sqrt{p_0} \braket{p_j \cos(\Delta \alpha_j)}_N \right) \Bigr] \textbf{M}_{j}.
 \label{eq:GIimageExpanded}
\end{align}

We see that the first term can be computed from the real part of Eq. \ref{eq:ObjMatrix}
\begin{align}
\frac{ \sqrt{p_0}}{N} \text{Re} \left[ \textbf{O} \right] :=& \frac{1}{N} \sum_{j=0}^{N-1} \sqrt{p_0 p_j} \cos(\Delta \alpha_j) \textbf{M}_j  \nonumber \\
=& \frac{\sqrt{p_0}}{ N} |\textbf{O}|\cos \left( \text{arg} \left( \textbf{O} \right) -\alpha_0 \right)
.
 \label{eq:term1}
\end{align}
For the second term in Eq.~\ref{eq:GIimageExpanded} we invoke Eq.~\ref{eq:Probmatrix} and compute,
\begin{align}
 \sum_{j=0}^{N-1} p_j \textbf{M}_{j} =&  \frac{1}{2  N } \sum_{j=0}^{N-1} p_j \textbf{H}_{d} \textbf{R}_{j} \textbf{H}_{d}^\dagger \nonumber \\
=&  \frac{1}{2 N } \textbf{H}_{d} \left( \sum_{j=0}^{N-1} p_j  \textbf{R}_{j} \right) \textbf{H}_{d}^\dagger  \nonumber \\
=&  \frac{1}{2 N } \textbf{H}_{d} \textbf{P} \textbf{H}_{d}^\dagger,
\label{eq:term2}
\end{align}
showing that it can be computed from a simple Hadamard transform of the probability matrix.
Furthermore, the matrix $\textbf{P}$ can also be computed via the Hadamard transform, as $P = |\textbf{H}_{d} \textbf{O} \textbf{H}_{d}|^2$, where $|\cdot|^2$ is the element wise absolute value squared.  This can be seen by substituting Eq.~\ref{eq:ObjMatrix} as follows,
\begin{align}
\textbf{P} = |\textbf{H}_{d} \textbf{O} \textbf{H}_{d}|^2 =& | \textbf{H}_{d}  \sum_{k}^{N} \sqrt{p_k} e^{i\alpha_k} \textbf{M}_k \textbf{H}_{d}^\dagger |^2, \nonumber \\
=&  \sum_{k}^{N} | \sqrt{p_k} e^{i\alpha_k}|^2 \textbf{H}_{d}  \textbf{M}_k \textbf{H}_{d}^\dagger, \nonumber \\
=&  \left( \sum_{k=0}^{N-1} p_k  \textbf{R}_{k} \right). \nonumber \\
\end{align}
Computing the matrix $\textbf{P}$  using the transform is less computationally expensive than computing the coefficients $p_k$ via the modal overlaps.

\begin{figure*}[t]
\includegraphics[width=\linewidth]{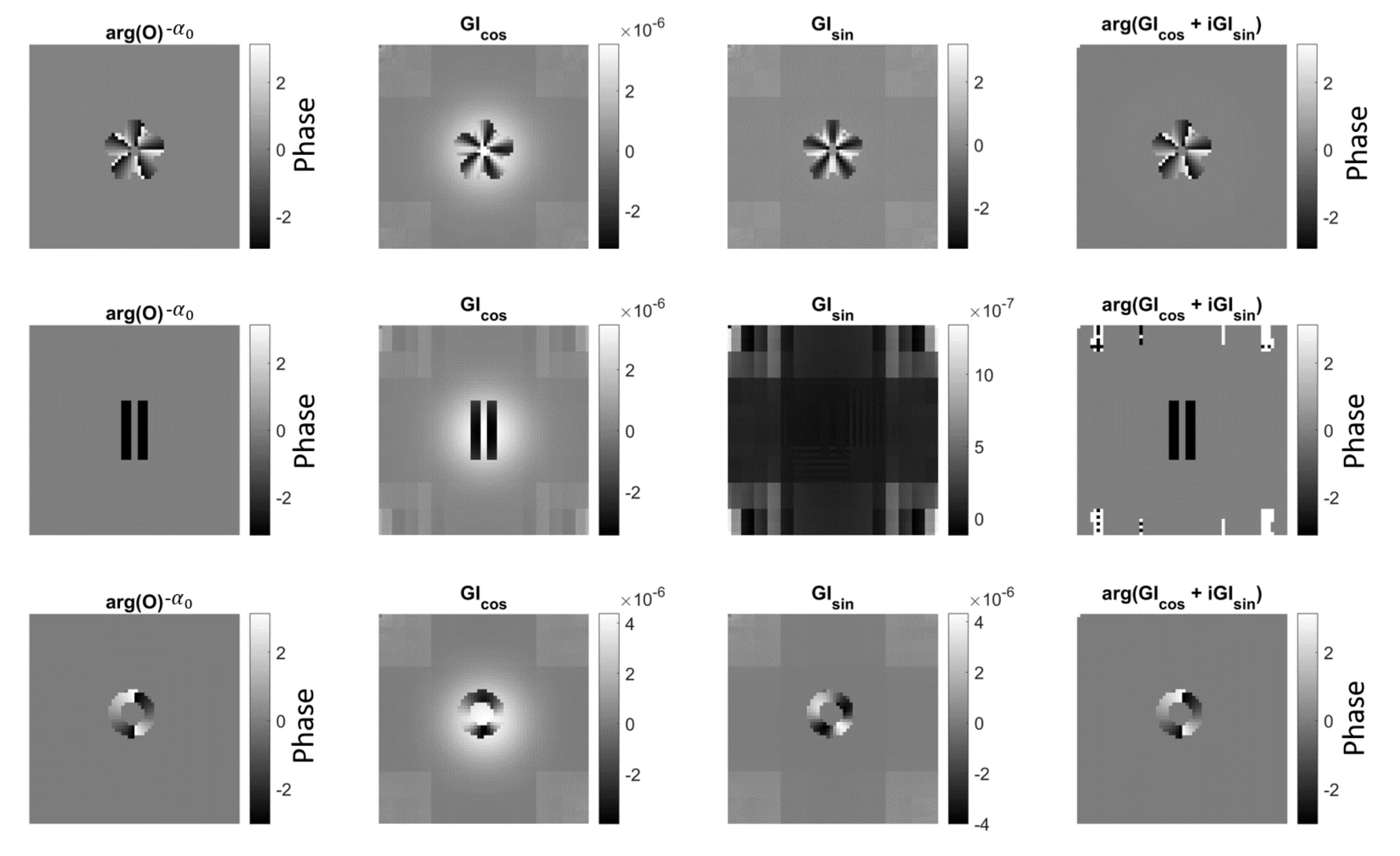}
\caption{ Example theoretical phase reconstruction from the ghost images.}
\centering
\label{fig:PhaseRecon1}
\end{figure*}

\begin{figure*}[b]
\includegraphics[width=\linewidth]{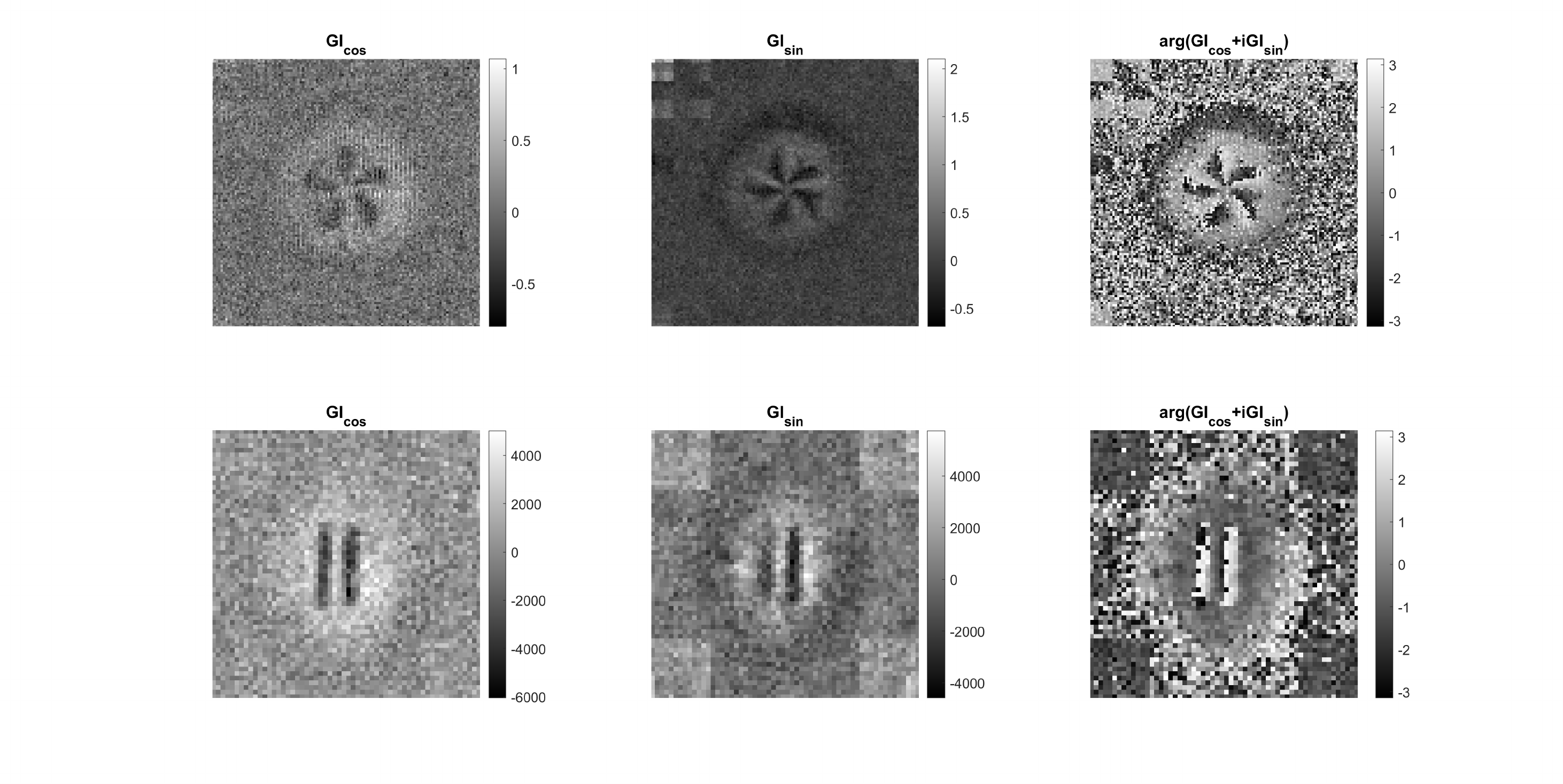}
\caption{ Uncropped experimentally measured ghost images.}
\centering
\label{fig:PhaseRecon2}
\end{figure*}

Finally, as for the last term in Eq.~\ref{eq:GIimageExpanded}, we can take the ensemble averages $g_{\cos} = \sqrt{N}/N \left( \braket{p_j}_N/2 + \sqrt{p_0} \braket{p_j \cos(\Delta \alpha_j)}_N \right)$  outside the summation and remain with $\sum_{j=0}^{N-1} 1/\sqrt{N} \textbf{M}_{j} = \textbf{R}_0$ which is the DC component.

Using Eq.~\ref{eq:term1} and Eq.~\ref{eq:term2} we can write Eq.~\ref{eq:GIimageExpanded} as 
\begin{align}
 \textbf{GI}_{\cos} =&   \frac{ 1}{N} \left( \sqrt{p_0} \text{Re} \left( \textbf{O} \right) +  \frac{1}{2} \textbf{H}_{d} \textbf{P} \textbf{H}_{d}^\dagger - g_{\cos} \textbf{R}_0 \right).
 \label{eq:GIimageSolved2}
\end{align}

We summarise the meaning of each term: (i) term 1 is the real part of the object profile up a global phase $\alpha_0$, term 2 depends on the spectral density of the object, while the last term corresponds to the DC component.

Following the same arguments, the imaginary part of the ghost image can be computed by applying the same analysis but with $\textbf{T}_j= \frac{1}{\sqrt{2}} ( \textbf{M}_j + i\textbf{M}_0)$, resulting in the probabilities becoming \\ $|c_j|^2= p_0/2 + p_j + \sqrt{p_o p_j}\sin(\Delta \alpha_j )$. Accordingly, the ghost image becomes
\begin{align}
 \textbf{GI}_{\sin} =&   \frac{1}{N} \left( \sqrt{p_0} \text{Im} \left( \textbf{O} \right) +  \frac{1}{2} \textbf{H}_{d} \textbf{P} \textbf{H}_{d}^\dagger - g_{\sin} \textbf{R}_0 \right),
 \label{eq:GIimageSolved3}
\end{align}
having an embedded imaginary part of the object profile  $\text{Im} \left( \textbf{O} \right) =|\textbf{O}|\sin \left( \text{arg} \left( \textbf{O} \right) -\alpha_0 \right)$ and $g_{\sin} = \sqrt{N}/N \left(  \braket{p_j}_N/2 + \sqrt{p_0} \braket{p_j \sin(\Delta \alpha_j)}_N \right)$

\bibliography{apssamp}

\end{document}